\documentclass[pra,showpacs,eqsecnum,twocolumn,amsmath,amssymb,nofootinbib]{revtex4}
\usepackage[swedish, british]{babel}
\usepackage{graphicx}
\usepackage{bm}
\newcommand{\beq}{\begin{equation}}
\newcommand{\eeq}{\end{equation}}
\newcommand{\beqa}{\begin{eqnarray}}
\newcommand{\eeqa}{\end{eqnarray}}
\newcommand{\ket}[1]{\lvert #1 \rangle}
\newcommand{\bra}[1]{\langle #1 |}
\newcommand{\opaa}{\hat{A}}
\newcommand{\opa}{\hat{a}}
\newcommand{\opad}{\hat{a}^\dagger}
\newcommand{\opaone}{\hat{a}_1}
\newcommand{\opatwo}{\hat{a}_2}
\newcommand{\opn}{\hat{n}}
\newcommand{\opnone}{\hat{n}_1}
\newcommand{\opntwo}{\hat{n}_2}
\newcommand{\opx}{\hat{x}}
\newcommand{\opspinn}{\hat{S_r}}
\newcommand{\tsingle}{\theta_\text{sing}}
\newcommand{\tsuper}{\theta_\text{sup}}
\newcommand{\abs}[1]{\lvert#1\rvert}
\newcommand{\ex}[1]{\langle#1\rangle}
\newcommand{\newtext}[1]{#1}

 \providecommand{\href}[2]{#2}
 \providecommand{\eprint}[2][http://arxiv.org/abs/]{\href{#1#2}{\texttt{#2}}}
 \newcommand{\arxiveprint}{\eprint}

\begin{document}

\title{A size criterion for macroscopic superposition states}

\author{Gunnar Bj\"{o}rk}
\email{gunnarb@imit.kth.se} \homepage{http://www.ele.kth.se/QEO/}
\affiliation{Department of Microelectronics and Information
Technology, Royal Institute of Technology (KTH), Electrum 229,
SE-164 40 Kista, Sweden}

\author{Piero~G.~Luca Mana}
\email{mana@imit.kth.se} \affiliation{Department of
Microelectronics and Information Technology, Royal Institute of
Technology (KTH), Electrum 229, SE-164 40 Kista, Sweden}

\date{9 June 2004}

\begin{abstract}
An operational measure to quantify the sizes of some
``macroscopic quantum superpositions'', realized in recent
experiments, is proposed. The measure is based on the fact
that a superposition presents greater sensitivity in
interferometric applications than its superposed
constituent states. This enhanced sensitivity, or
``interference utility'', may then be used as a size
criterion among superpositions.
\end{abstract}


\maketitle

\section{Introduction}
Soon after the birth of quantum mechanics it was noticed
that the new theory suggests the existence of new
statistical properties in peculiar experimental
arrangements~\cite{Einsteinetal1935,Bohr1935,Schroedinger1935}.
This gave rise to a discussion on interpretative issues
which lasts still today and which has been accompanied by
experimental efforts to study these peculiar statistical
features. Some of the peculiarities seem to originate when
we append the adjective \emph{macroscopic} to the concept
of \emph{superposition}.

A superposition of two given states is, very simply, a
third, different state having some precise statistical
relationships with the former. However, it is more often
sadly (re)presented as a sort of ``coexistence'' of the two
original states, and since these can have quite different
and incompatible characteristics, their ``coexistence''
assumes then a mysterious and counterintuitive or
paradoxical nature. This can be tolerated when the states
refer to imperceptible and ephemeral objects, but is
unacceptable if the latter are not so ephemeral, or even
visible, like the cat in Schr\"odinger's
example~\cite{Schroedinger1935} --- here enters the
adjective \emph{macroscopic}.

To us, this adjective implies two things, \newtext{which
are also crucial points in Schr\"odinger's example}.
Firstly, the system under study should be highly excited,
or consist of a large number of quantized subsystems
\newtext{(the ``cat'')}. Ideally, it should be possible to
infer the system's state through classical meter such as
the eye, a ruler, a photographic film, a voltmeter, a
magnetometer etc. Secondly, each of the two states the
superposition consists of, should be in an as classical
state as quantum mechanics allow \newtext{(the ``alive''
and ``dead'' states)}. Such states are often referred to as
semiclassical, and examples include coherent states, spin
coherent states or a collection of particles whose
deviation from their mean positions are uncorrelated.
Operationally, this implies that the two constituent states
should not be eigenstates of the measured observable as
this is unnatural and atypical for non-ephemeral objects.
(Moreover, superpositions of highly excited eigenstates are
difficult to prepare and therefore they are not the
experimentalist's choice of macroscopic superpositions, as
shown below.)

Different interpretation schools face the issue of
macroscopic superpositions in different ways, but it is not
the purpose of this paper to expose, or to further, these
interpretations. It may just be said that the situation is
fortunately not so
paradoxical~\cite{Peresetal1964,Peres1980}. The results
achieved in noteworthy experimental
efforts~\cite{Awschalometal1990,Awschalometal1992,Ciracetal1998,%
Ruostekoskietal1998,Arndtetal1999,Harrisetal1999,Nakamuraetal1999,%
Friedmanetal2000,Vanderwaletal2000,Bonvilleetal2001,Julsgaardetal2001,%
Marshalletal2002,Auffevesetal2003,Everittetal2003,%
Nairzetal2003,Ghoshetal2003} (see
Ref.~\onlinecite{Leggett2002b} for a recent review) in the
quest of creating superpositions as ``macroscopic'' as
possible, can be seen as a test of our interpretation(s)
and understanding of this issue. The variety and difference
of the phenomena studied in these technically challenging
experiments, which range from heavy-molecule interferometry
to superconducting devices, is fascinating, but it is also
the source of difficulties in estimating which of the
superpositions hitherto realized is the most
``macroscopic'', and, \emph{quantitatively}, how
macroscopic it is. Some measures have been proposed for
this purpose by Leggett~\cite{Leggett1980,Leggett2002b}
and, recently, by D\"ur et al.~\cite{Dueretal2002}.

Leggett's measure, called
`disconnectivity'~\cite{Leggett1980}, has some affinity
with entanglement measures, and shares a non-operational
nature with many of them. The idea behind it is to count
the ``effective'' number of particles involved in a
superposition, or more precisely, the effective number of
quantum-correlated particles. This is achieved, roughly
speaking, by checking how large the entropies of all
possible reduced states are. The disconnectivity is not
invariant under global unitary transformations, which is
equivalent to stating that there is a ``preferred'' set of
basis states. These are related to the (only) states which
are usually observed at the macroscopic level.

The measure proposed by D\"ur et al.~\cite{Dueretal2002},
instead, is based on the study of the similarities between
a superposition of the form
$2^{-1/2}(\ket{\phi_+}^{\otimes N} + \ket{\phi_-}^{\otimes
N})$, where $|\langle \phi_- \ket{\phi_+}| \neq 0$, and a
standard state of the form $2^{-1/2} (\ket{0}^{\otimes n}
+\ket{1}^{\otimes n})$, assessing for which $n$ they are
most similar. Similarity is established, roughly speaking,
by comparing either decoherence times or entanglement
resources; notably these two criteria lead to the same
result. Also in this case the measure depends on a choice
of a \emph{preferred} set of basis states.

A characteristic common to these two measures is that they
emphasize the r\^ole of the number of particles, or modes,
participating in the superposition; in a sense, they
associate the adjective `macroscopic' only with these
features. However, though the number of particles is
certainly an important factor, one may asks whether it is
the only one that should be associated to the word
`macroscopic'. Above, we have stated that we do not believe
so. To clarify and justify the reason for this, consider a
standard two-arm interferometer, and the two states in
which we have ten photons (with frequency in the visible
spectrum) in one arm or, respectively, in the other. Such
states could be distinguished by the naked
eye~\cite{Hecht1942,Bayloretal1979}, and hence should be
rightly called `macroscopic', even though they engage only
two modes (but ten quanta) Their superposition should thus,
in our opinion, be a `macroscopic superposition' but the
disconnectivity of this state is only $2$, and the
disconnectivity is independent of the state's excitation
$N$.

Indeed, besides disconnectivity, Leggett has proposed
another, ``auxiliary'' measure, the `extensive
difference'~\cite{Leggett2002b}, whose idea has something
in common with the preceding remark. Roughly speaking, it
is the difference between the expectation values for the
superposed states of a particular (``extensive'')
observable, the latter being that for which this difference
is maximal; the expectation values are of course expressed
with respect to some ``atomic'' unit.

\section{The general idea: Interference utility}
\label{sec:idea}

In line with the last argument, we propose an operationally
defined measure based on the properties of a preferred
observable, rather than on the definition of an effective
number of particles, or modes. The idea is based on the
following points:

First, one tries to identify a \emph{`preferred'
observable} which is especially related to the experiment
in question. The fact that such an observable must indeed
be present is strongly suggested by the existence of
preferred basis states, as noted above: \emph{without such
a set of states there would be no point in emphasizing
superpositions, since every state can always be written as
some superposition, and every pure superposition as a
single state}. Note that the adjective `preferred' is not
meant here in some vague sense of ``preferred by nature'',
or in a sense analogous to Zurek's ``pointer observable and
states''~\citep{Zurek2003}; instead, it is meant to denote
the observable which is most useful for a physicist in a
particular (interferometric) application, and hence depend
on the experimental and research context.

Second, one considers two classical, macroscopic states
that can be separately characterized by a more or less
peaked distribution in the eigenvalue spectrum of the
preferred observable, centered around a given eigenvalue.
This distribution has a breadth, which is usually implied
by the semiclassical nature of the state, and implies the
latter's usefulness in an interferometric application with
the preferred observable. If the probability distribution
of each state is a Dirac delta the state is typically no
longer classical, or macroscopic, or interferometrically
useful in our sense. A semiclassical state typically has
an eigenvalue spectrum with a width proportional to the
square root of its excitation or constituent particles.

Third, one imagines to make the superposition state evolve
under the operator generated by the preferred observable.
If the state is a superposition, this evolution will
produce a rapid oscillation between the state and an
orthogonal state (i.e., it will give rise to interference).
The oscillation frequency is higher than for the single
superposed semiclassical states (due to their breadths),
and it would not be faster if the state were just a
mixture. The crucial point here is that the larger is the
eigenvalue separation between the two states, the higher is
this oscillation frequency. Conversely, in an application
based on such an interference, the oscillation frequency is
an indication of the state's sensitivity as a probe, or its
\emph{utility} in an interferometric application, with
respect to the utility of the semiclassical states.

Hence, to appreciate if and how much this oscillation is
due to the macroscopic quantum superposition, one compares
it with the frequency of the oscillation due to the spread
of the wave-function of each of the two states in the
superposition, i.e., one compares the oscillation frequency
of the superposed state with that of the single states
(which is equivalent to that of the corresponding mixture).

Let us remark, together with Leggett~\cite{Leggett2002},
that the issue about the word `macroscopic' presents many
subjective elements --- which are unavoidable.  The purpose
of this paper is thus not to eliminate these elements, but
to try to frame them in a more operational context: they
will enter in the choice of the `preferred observable'
discussed above, but such a choice can be better motivated
on, e.g., applicational grounds. Here also enter in the
interpretation of the adjective `macroscopic'.


\section{Mathematical formulation}

\begin{figure}
\includegraphics[width=7cm,height=4cm]{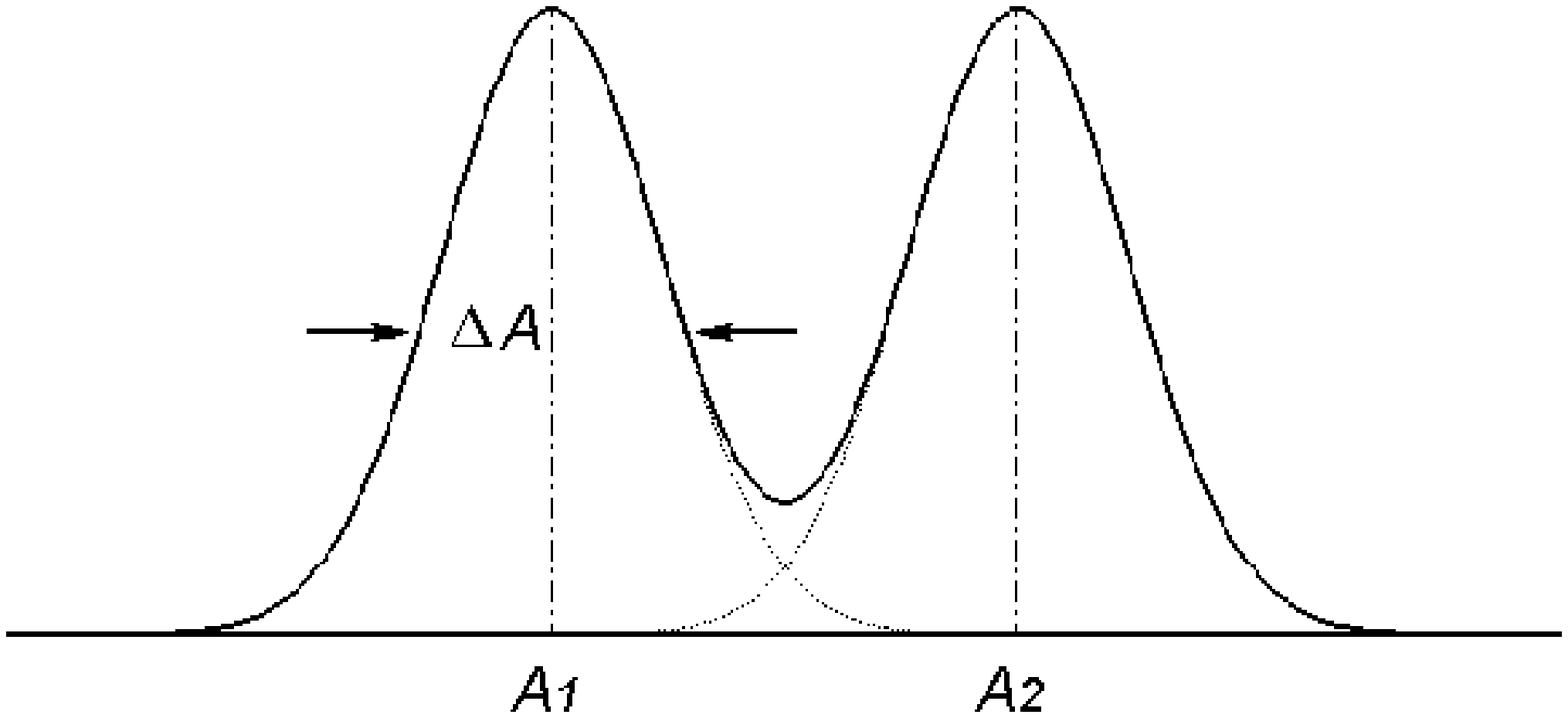}
\caption{A bimodal probability distribution $P(A) =
|f(A)|^2$ with peaks of width $\Delta A\equiv w$ centered
at the values $A_1$ and $A_2$.} \label{fig: distribution}
\end{figure}

A macroscopic superposition, in its simplest form, consists
of a (pure) state for which the probability distribution
for the preferred observable is bi-modal. (One can also
imagine cases where the distribution is multi-modal, but
this case will not be treated here.) For what follows, the
exact form of the distribution is inconsequential, as long
as it is reasonably smooth. To put what was just said in
more concrete dress, consider a Young double slit
experiment. The pertinent preferred operator in this case
is the position operator $\opx$. A particle incident on the
double slit will have a bimodal probability distribution
$P(x) = |\bra{x}\psi\rangle|^2$ at the exit side of the
slit. (Here, it is assumed that origin of the bimodal
probability distribution is a bimodal probability amplitude
distribution. That is, we have a pure superposition state
and not a mixed state.) Hence, if we expand the
superposition state in the (complete) set of eigenvectors
$\{\ket{A}|A\in\mathbb{R}\}$ of the preferred operator
$\opaa$, we will get \beq
\label{eq:stateexpanded}\ket{\psi} = \int_\mathbb{R}
\bra{A}\psi\rangle \ket{A} dA = \int_\mathbb{R} f(A)
\ket{A} dA . \eeq As mentioned above, we assume that
$|f(A)|^2$ deviates appreciably from zero only within an
interval around two values of $\opaa$ that we will denote
$A_1$ and $A_2$ (see Fig.~\ref{fig: distribution}), hence
it can be expressed as
\begin{equation}
\label{eq:bimodalform} \lvert f(A)\rvert^2 = c(A-A_1) +c(A-A_2),
\end{equation}
where $c(A)$ is a (positive and appropriately normalized)
roughly even function with a single maximum at the origin
$A=0$.

A binary, equal, superposition has the property that it
evolves into an orthogonal, or almost orthogonal state,
upon a relative phase-shift of $\pi$ between its two
components. To achieve such a phase-shift, we evolve the
state~(\ref{eq:stateexpanded}) under the action of its
preferred operator, i.e., under the unitary evolution
operator $\exp(i \theta \opaa)$, where the real parameter
$\theta$ characterizes the interaction time or strength.
The evolved state becomes \beq \ket{\psi(\theta)} = e^{i
\theta \opaa} \ket{\psi} = \int_\mathbb{R} e^{i \theta A}
f(A) \ket{A} dA,\label{eq: evolved state} \eeq and its
inner product with the original state is
\begin{equation}
\label{eq:interference1}
\begin{split}
\bra{\psi} e^{i \theta \opaa} \ket{\psi} &=
\int_\mathbb{R} e^{i\theta A} \lvert f(A)\rvert^2\, dA \\
&=
\int_\mathbb{R} e^{i \theta A}  c(A-A_1) \,dA \\
&\quad+
\int_\mathbb{R} e^{i \theta A}  c(A-A_2) \,dA, \\
&= (e^{i \theta A_1}+e^{i \theta A_2})
 \int_\mathbb{R} e^{i \theta A}  c(A) \,dA \\
\end{split}
\end{equation}
where we have used the bi-modal
assumption~\eqref{eq:bimodalform}. The modulus of this
inner product can thus be written
\begin{equation}
\label{eq:interference_expr} \lvert \bra{\psi} e^{i \theta \opaa}
\ket{\psi}\rvert = 2 \left\lvert \cos \frac{\theta(A_2 -A_1)}{2}
\int_{\mathbb{R}} e^{i \theta A} c(A) \,dA \right\rvert.
\end{equation}

If the function $c(a)$ is smooth, then we can deduce that
the integral in (\ref{eq:interference_expr}) will have its
first minimum when $\theta \approx
\tsingle\equiv\pi/\Delta A$, where $\Delta A$ is the width
(e.g., FWHM) of $c(A)$ (and hence the width each of the
two peaks of $\lvert f(A)\rvert^2$). However, the right
hand side of (\ref{eq:interference_expr}) also contains
the factor $\cos[\theta(A_2-A_1)/2]$. This interference
factor becomes zero when $\theta = \tsuper \equiv\pi/(A_2
- A_1)$ and appears only because the system is in a
superposition state. For large (``macroscopic'') values of
the separation between $A_1$ and $A_2$, the evolution into
an orthogonal state due to this factor may be
significantly faster than the evolution due to the width
$\Delta A$ of the probability distribution peaks.
Therefore, a measure of the system's
``\emph{interferometric macroscopality}'' $M$ is the
inverse ratio between the interaction needed to evolve the
system into an orthogonal state when it is in a
superposition, and when it is not. Thus, \beq M =
\frac{\tsingle}{\tsuper}\approx \frac{|A_1-A_2|}{\Delta A}
. \label{eq: macroscopality} \eeq Being a ratio, $M$ is
dimensionless. Moreover, it is independent of the
specifics of the measurement such as geometry, field
strengths, etc., as all these experimental parameters are
built into the evolution parameter $\theta$. That is, the
measure compares the evolution under identical
experimental conditions for a system and a binary
superposition of the very same system. Moreover, the
measure is applicable to all bimodal superpositions, it
is, i.e., easy to extend the analysis to observables with
a discrete eigenvalue spectrum as will be done in
Sec.~\ref{Number-state superpositions},
below.\footnote{The analysis just presented can of course
be also formulated in Wigner-function
formalism~\citep{Wigner1932,Cohen1966,Hilleryetal1984,Cohen1989,leonhardt1997};
the presence of ``semiclassical'' states perhaps suggest
that such a formalism would be more easily applied.
However, it is easy to realise that this is not the case:
first, we would have the additional spurious presence of
generalised phase-space observables, not necessarily
related with the preferred observable nor with the
superposition state; second, evolution should be computed
through a generalised Liouville equation; third, the
overlap between the various (evolved) states should be
computed through integrations (involving variables not
directly related to the problem). Hence, Dirac's formalism
is more suited to the mathematical expression of the above
ideas.} In contrast to disconnectivity, the measure does
not favor (nor disfavor) states with many particles or
modes. The measure just tells us how much the evolution of
the state will be accelerated by the means of the binary
superposition.

\section{Examples}

\subsection{Superpositions of number states}
\label{Number-state superpositions}

Let us start by examining a class of states for which the
measure $M$ is not directly applicable (and neither is the
measure proposed by D\"ur et al.). An example of such a
state is the two-mode, bosonic state \beq
(\ket{0}\otimes\ket{N} + \ket{N}\otimes\ket{0})/\sqrt{2} .
\label{eq: boson cat} \eeq Each of the terms in the
superposition state is a two-mode number-state.

\begin{figure}
\includegraphics[width=7cm]{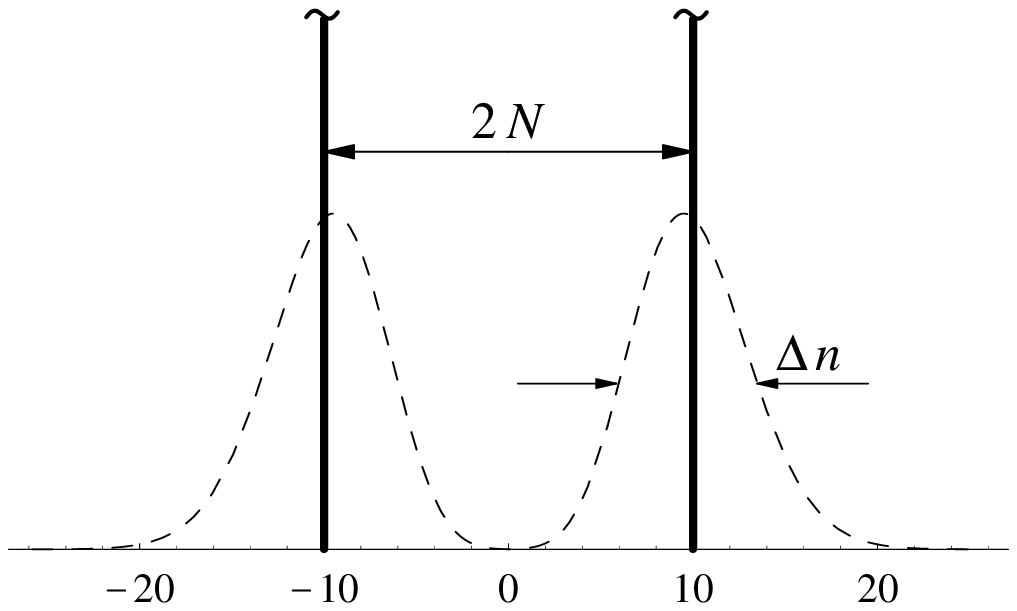}
\caption{Probability distributions for
the preferred observable $n=(\opnone - \opntwo)/2$. The
two (truncated) thick lines represent the
(Kronecker-delta-like) distribution for the state
$(\ket{0}\otimes\ket{N} + \ket{N}\otimes\ket{0})/\sqrt{2}$
with $N=20$, which becomes orthogonal for
$\tsingle=\pi/N$. The dashed distributions correspond to
coherent states $\ket{0}\otimes\ket{\alpha}$ and
$\ket{\alpha}\otimes\ket{0}$ with
$\abs{\alpha}^2=\abs{\ex{n}}=20$ and a dispersion $\Delta
n\approx\sqrt{20}$, which become almost orthogonal for
$\tsuper\approx\pi/\sqrt{N}$. (Note that these
distributions are discrete; their envelopes have been used
here for simplicity.)}\label{fig:numberst}
\end{figure}

Since both components of the superposition state lack
dispersion for the preferred observable they do not evolve
under the preferred evolution operator $\exp[i \theta
(\opnone - \opntwo)/2]$, where $\opnone$ ($\opntwo$)is the
number operator operating on the left (right) mode in the
tensor product. One can ascribe this ``nonevolution'' to
the fact that the number states are eigenstates to our
preferred observable. Moreover, the number states are
highly nonclassical (and consequently very fragile with
respect to dissipation). Hence, although the state is in a
highly excited superposition when $N$ is large, it is not
in a superposition of two semiclassical states, and
therefore, it is not in the spirit of Schr\"odinger's
example. In this case, we can still derive an
interferometric macroscopality by approximating each of the
superposition constituent states with a semiclassical state
(cf.\ Fig.~\ref{fig:numberst}). E.g., we can compare the
evolution of the superposition state with that of a
coherent state having a photon number expectation value of
$N$. Since the dispersion $\langle(\opn - \langle \opn
\rangle)^2 \rangle$ of a coherent state is $\langle \opn
\rangle$, we find that for such a state we get $\Delta n =
\sqrt{N}$. (This result is identical to that we would have
obtained if we instead had considered a two-mode coherent
state.) The state in Eq. (\ref{eq: boson cat}) becomes
orthogonal for $\tsingle=\pi/N$. Hence, the interferometric
macroscopality of the state becomes \beq
M\approx\frac{\pi/\sqrt{N}}{\pi/N}\approx \sqrt{N} . \eeq
(See Fig.~\ref{fig:numberst}.) This result makes intuitive
sense. In interferometry, this $\sqrt{N}$ improvement in
utility denoted the difference between the standard quantum
limit and the Heisenberg
limit~\cite{Caves1980,Caves1981,Yurkeetal1986,Grangieretal1987,Fabreetal2000,Trepsetal2002}.

The scaling would not change qualitatively if we considered
a coherent state with mean photon number $N/2$, nor would
it change if we had considered a slightly squeezed coherent
state (with a constant squeezing
parameter)\footnote{Imagining to have the technology to
create squeezed states of any degree of squeezing, and to
adapt the squeezing parameter for any choice $N$, the
scaling would become $N^{2/3}$~\citep{Loudonetal1987};
however, the comparison with squeezed states would not be
in the spirit of the present paper's ideas, as squeezed
states are not semiclassical~\citep{leonhardt1997}.}

The scaling $\sqrt{N}$ for the interferometric utility is
to be compared with the fragility of the state under
decoherence, which scales as $N$. This means, intuitively,
that the ``cost'' for preserving this kind of state
increases faster (by a factor $\approx\sqrt{N}$) than the
state's sensitivity in interferometric application does.

This state also is a good example for the strange behavior
of the disconnectivity as a measure of how macroscopic the
superposition above is, for the disconnectivity is always
unity for the state, irrespective of $N$. The simple
explanation is that we do not increase the ``size'' of the
state by increasing the number of modes, but instead
increase only the number of excitations.

\subsection{Qubit states}
\label{Qubit states}

A simple state that has been used as a model system to
quantify the size of macroscopic superpositions is the
$N$-qubit state \beq \frac{1}{\sqrt{2}}
(\ket{\phi_+}^{\otimes N} + \ket{\phi_-}^{\otimes N}) ,
\label{eq: many qubits} \eeq where \beq \ket{\phi_\pm} =
\cos(\varphi \pm \epsilon/2)\,\ket{1/2} + \sin(\varphi \pm
\epsilon/2)\,\ket{-1/2}, \label{eq: one qubit} \eeq where,
without loss of generality, we have taken the two qubit
basis states to be spin 1/2 eigenstates along some axis. We
note that the overlap $|\langle \phi_-\ket{\phi_+}| =
|\cos\epsilon|$. Hence, for $0<\epsilon \ll 1$, we have
$|\langle \phi_-\ket{\phi_+}|^2 \approx 1 - \epsilon^2,$
and the qubit states have a large overlap. However, as \beq
|\bra{ \phi_-}^{\otimes N} \ket{\phi_+}^{\otimes N}| =
|\cos^N\epsilon| , \eeq the many-qubit state has almost
vanishing overlap as soon as $N \gg 1$ and $\epsilon>
1/\sqrt{N}$. To be more specific, for $\epsilon =
\arccos(e^{-1/N}) \approx \sqrt{2/N}$, the overlap is
$1/e$. Hence, the $N$-qubit states $\ket{\phi_\pm}^{\otimes
N}$ are almost orthogonal for $\epsilon > 1/\sqrt{N}$.

In order to calculate the interferometric macroscopality of
the $N$-qubit state, we choose as preferred observable the
total spin $\opspinn=\sum_{n=1}^N \hat{\sigma}_{n,r}$ (in
units of $\hbar$) along the axis $r$ defined by the basis
states. We compute the mean \beq\label{eq:mean-qubit}
\bra{\phi_\pm}^{\otimes N} \opspinn \ket{\phi_\pm}^{\otimes
N} = N \cos(2 \varphi \pm \epsilon)/2 , \eeq and the
variance \beq\label{eq:variance-qubit}
\bra{\phi_\pm}^{\otimes N} \Delta \opspinn^2
\ket{\phi_\pm}^{\otimes N} = N [1-\cos^2(2 \varphi \pm
\epsilon)]/4 . \eeq We see that in order to make the
difference between the expectation values of the two
superposed states $\ket{\phi_+}^{\otimes N}$ and
$\ket{\phi_-}^{\otimes N}$ as large as possible, we should
choose $\varphi=\pi/4$, i.e., the preferred observable
should be along an axis forming an angle of $\pi/2$ with
the superposition state's average spin, in a Bloch-sphere
representation. For this choice, $|\bra{ \phi_+} \opspinn
\ket{ \phi_+} - \bra{ \phi_-} \opspinn \ket{ \phi_-}|=N
|\sin \epsilon | \approx N \epsilon$ and the width of each
spin probability distribution (the square root of each
distribution's variance) is approximately $\sqrt{N}/2$. We
immediately see that as soon as $\epsilon > 1/\sqrt{N}$, we
get a bimodal distribution for the superposed state. This
fact is supported by the condition derived previously for
the two states to have a small overlap. Thus for the
state~(\ref{eq: many qubits}) we get \beq M \approx \frac{N
\epsilon}{\sqrt{N}/2}= 2 \sqrt{N} \epsilon . \eeq This
measure can be compared to the measure by D\"ur et al.,
where they conclude that the state's macroscopic size is $N
\epsilon^2$. However, their measure is based on the premise
that the macroscopic size of the state $2^{-1/2}
(\ket{0}^{\otimes n} +\ket{1}^{\otimes n})$ is $n$. In
contrast, we find that for this state $M =
n/\sqrt{n}=\sqrt{n}$. Because $\bra{0}^{\otimes n} \Delta
\opspinn^2\ket{0}^{\otimes n} = \bra{1}^{\otimes n} \Delta
\opspinn^2\ket{1}^{\otimes n} = 0$, we have used the method
described in the previous subsection, comparing the
evolution time of the superposition state with that of a
spin-coherent state with mean excitation $n$. We see that
our measure is proportional to the square root of the
measure of D\"ur et al., when a comparison is applicable.
The difference between the measures can be traced to the
operational questions they answer. In our case it is the
state's interference utility, in the case of D\"ur et al.,
it is how many qubits a GHZ state distilled from the
macroscopic superposition contains.

Considering again Eqs.~\eqref{eq:mean-qubit}
and~\eqref{eq:variance-qubit}, we see that a choice of an
observable forming a vanishing angle with the superposition
state's average spin would have led to an interference
utility $M=0$: with respect to such observable, the
state~\eqref{eq: many qubits} would present very small ---
i.e., non-macroscopic --- interference effects.

It is also worth noticing that the disconnectivity of the
state~(\ref{eq: many qubits}) is $N$, irrespective of
$\epsilon$, as long as $\epsilon \neq 0$. The
disconnectivity vanishes only when $\epsilon=0$ exactly,
even if for $\epsilon < 1/\sqrt{N}$ the state does not
represent a macroscopic superposition any longer. The
disconnectivity's invariance with respect to $\epsilon$
clearly demonstrates the measure's exclusive dependence on
the particle (or mode) number and its ignorance of the
state of the particles (modes).

\subsection{A superposition of coherent states}
\label{coherent superpos}

Several authors have suggested to use a binary
superposition of coherent states to make macroscopic
superpositions. At least one experiment has been performed
on such a state, that is of the form \beq \ket{\psi} =
\frac{1}{\sqrt{N}} (\ket{ e^{i \varphi}|\alpha|} + \ket{
e^{-i \varphi}|\alpha|}) , \label{coherent SC} \eeq where
the normalization factor \beq N=2\{1 + \exp[-|\alpha|^2
(1-\cos 2 \varphi)]\cos(|\alpha|^2 \sin 2 \varphi)\} \eeq
and where $\opa \ket{ e^{i \varphi}|\alpha|} = |\alpha|
e^{i \varphi} \ket{ e^{i \varphi}|\alpha|}$. If such a
state's quadrature-amplitude distribution is measured,
where the Hermitian quadrature amplitude operator is
defined $\opatwo=(\opa - \opad)/2 i$, a bimodal
distribution will be found, provided that the
superposition ``distance'' $D = 2|\alpha| \sin \varphi$ is
sufficiently large. Hence, the preferred evolution
operator is $\exp(i \theta \opatwo)$. Let us first see how
a coherent state evolves towards orthogonality under this
operator. The dispersion of $\opatwo$ of the coherent
state is $1/2$. Hence, we can expect the coherent state to
evolve into an (almost) orthogonal after an interaction of
$\theta = 2 \pi$.

In a more detailed calculation, we use the fact that
\begin{equation}
\begin{split}
e^{i \theta \opatwo} = e^{(- \theta \opad + \theta \opa)/2} &=
e^{- \theta \opad/2} e^{\theta \opa/2} e^{-[-\theta \opad,\theta
\opa]/8}\\
&= e^{-\theta^2/8}e^{- \theta \opad/2} e^{\theta \opa/2} .
\end{split}
\end{equation}
Since coherent states are eigenstates to the operator
$\opa$, it is easy to compute \beq \bra{\alpha}e^{i \theta
\opatwo}\ket{\alpha} = e^{-\theta^2/8}e^{i \theta
\text{Im}\{\alpha\}} , \eeq where Im$\{z\}$ denotes the
imaginary part of $z$ and where we have assumed that
$|\alpha| \neq 0$. We see that the magnitude of the overlap
between a coherent state and an evolved copy of itself
decreases with time as $e^{-\theta^2/8}$. Note that the
result is independent of $|\alpha|$. The evolved state
never becomes fully orthogonal to the unevolved state, the
overlap goes only asymptotically toward zero, but we can
define $\tsingle\approx2 \sqrt{2}$ as the typical
``orthogonality'' time. (The crude calculation above
yielded $\tsingle = 2 \pi$, roughly a factor of two
higher.) In this time the (absolute value of the) overlap
has decreased from unity to $1/e$. Applying the same
evolution operator to the state $\ket{\psi}$, above, one
gets
\begin{equation}
\begin{split}
\bra{\psi}e^{i \theta \opatwo}\ket{\psi}  &=  \frac{2
e^{-\theta^2/8}}{N}[\cos(\theta |\alpha| \sin \varphi)
\\
&\quad+ e^{-|\alpha|^2(1-\cos 2 \varphi)} \cos(|\alpha|^2 \sin 2
\varphi)] .
\end{split}
\end{equation}
In this case, for reasonable large values of $|\alpha|$ and
$\varphi$ (when the rightmost term of the equation above is
negligible) the state becomes almost orthogonal under
evolution, and this first happens when \beq \theta =
\tsuper=\frac{\pi}{2 |\alpha| \sin \varphi} . \eeq Hence,
the superposition's interferometric macroscopality will be
\beq M =\frac{\tsingle}{\tsuper}\approx \frac{2
\sqrt{2}}{\pi/2 |\alpha| \sin \varphi} \approx 2 |\alpha|
\sin \varphi. \eeq 
\begin{figure}
\includegraphics[width=7cm]{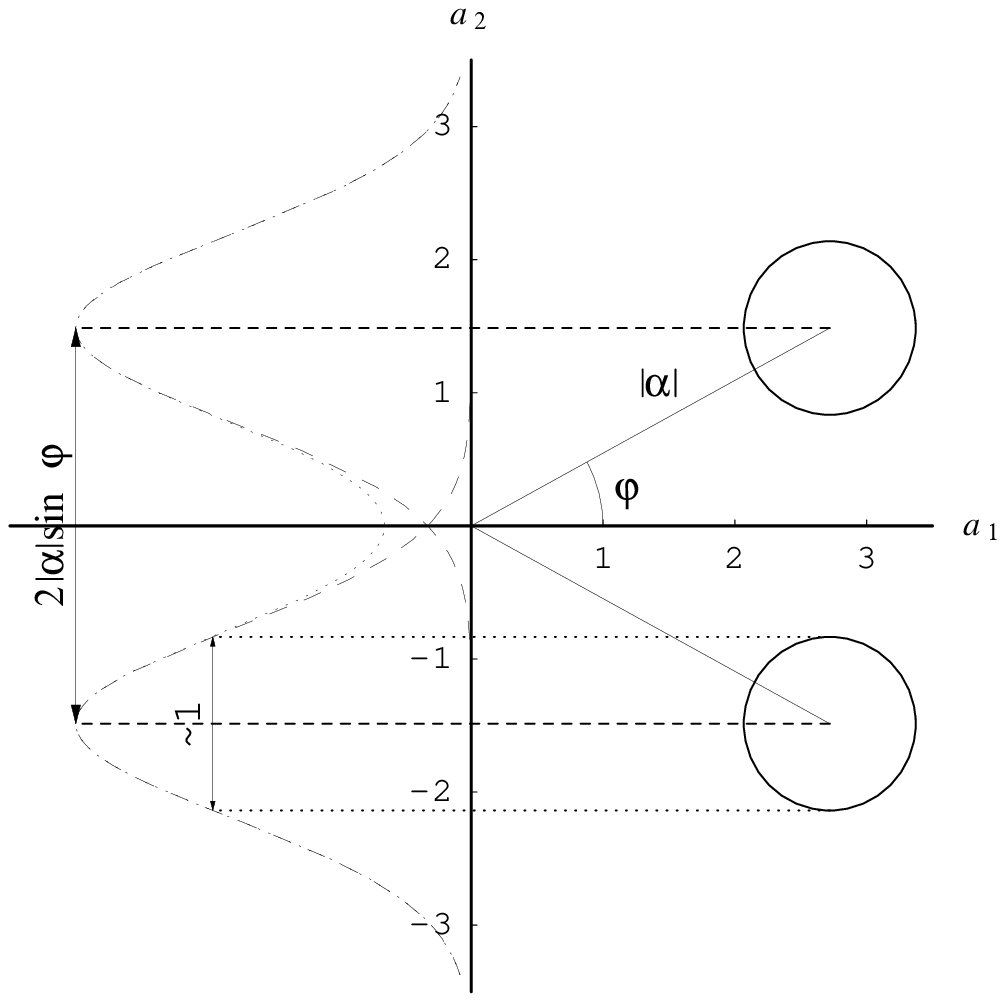}
\caption{Wigner-function representation of the coherent
states $\ket{ e^{i \varphi}\abs{\alpha}}$ and $\ket{ e^{-i
\varphi}\abs{\alpha}}$ (with $\abs{\alpha}=3.1$ and
$\varphi=0.5$) on the plane $\opaone\opatwo$. The
respective probability distributions for the preferred
observable $\opatwo$ are plotted on the left of the
latter's axis. The distance between the two distributions'
peaks is $2 |\alpha| \sin \varphi$, corresponding to a
$\tsuper=\tfrac{\pi}{2 |\alpha| \sin \varphi}$ for the
superposed state to become orthogonal; the widths of the
distributions is $\sim 1$, corresponding to a
$\tsingle\approx2\sqrt{2}$ for each single costituent
state to become orthogonal.} \label{fig:coherent}
\end{figure}
((See Fig.~\ref{fig:coherent}.) In experiments performed by
Brune et al.~\cite{Bruneetal1996} on Rydberg states at
microwave frequencies, a $|\alpha|$ of 3.1 was achieved,
and $\varphi$ was varied between approximately $0.1 - 0.5$
rad. Hence, an interferometric macroscopality of $\approx
6.2 \sin 0.5 \approx 3 $ was achieved. However, at the
largest angles $\varphi$ measured, the state had already
decohered substantially, so the superposition state was no
longer a pure superposition. It is worth noticing that the
decoherence, manifested in the decrease of the Ramsey
interference fringes measured in the experiment, scales as
$\exp(-D^2/2)$. The decoherence time is hence proportional
to the interferometric macroscopality squared of the state.
This is the same result we found for a rather different
state in Sec. \ref{Qubit states}. The result suggest that
that the found relation between the decoherence and the
interferometric macroscopality of the state is quite
general. The result also indicates that it will be
difficult to reap the full utility of macroscopic
superposition states in applications, unless dissipation is
kept to a minimum.

\subsection{Molecule interferometry}
\label{subsec. molecule interferometry}

A series of experiments have been performed on diffraction
of molecules with increasing atomic weight. The effort
started in the 1930's with interference experiments with
H$_2$ (weight 2 in atomic mass units), but recently the
field has used increasingly larger molecules going from
He$_2$, Li$_2$, Na$_2$, K$_2$ and I$_2$ (with weights 8,
14, 46, 78, and 254 atomic mass units, respectively) to
substantially heavier molecules such as C$_{60}$, C$_{70}$
to C$_{60}$F$_{48}$ (with weights 720, 840 and 1632 atomic
mass units, respectively). In one of the recent experiments
of this type \cite{Arndtetal1999}, a collimated beam of
C$_{60}$ molecules was diffracted by a free-standing
SiN$_x$ grating consisting of 50 nm wide slits with a 100
nm period $D$. A tree-peaked diffraction pattern was
observed a distance $L=1.25$ m behind the grating by means
of a scanning photo-ionization stage followed by a ion
detection unit. The distance from the central peak and the
nodes on each side of it was about 12 $\mu$m.

To analyze the experiment in terms of macroscopic
superpositions, we note that the molecular beam intensity
is such that the molecules are diffracting one by one.
Hence, we need not invoke quantum mechanics to explain the
diffraction pattern, but we can simply model the experiment
in terms of first-order wave interference. We note that for
a single slit of width $d$, the molecule intensity in the
direction $\theta$ from the slit normal direction is
proportional to \beq I \propto \int_{-d/2}^{d/2} e^{i k x
\sin(\theta)} \approx \int_{-d/2}^{d/2} e^{i k x \theta} ,
\label{eq: molecular diffraction} \eeq where $x$ is the
position coordinate across the slit, $k$ is the molecule's
de Broglie wave vector, we have assumed a constant
molecular beam amplitude across the slit, and diffraction
close to the normal has been considered. We see that in a
slit diffraction experiment, the position $x$ is the
preferred observable and the parameter $\theta$ can be
interpreted as the diffraction angle. We can express $k$ in
the molecular beam velocity $v = 220$ ms$^{-1}$, and the
molecular weight $m = 720$ atomic mass units as $k = 2 \pi
m v/h$, where $h$ is Planck's constant. We find that
orthogonality occurs when $\theta=\tsingle=h/d m v$.
Experimentally, this means that the single slit diffraction
pattern has its first node at this angle, or at a distance
$\approx h L/d m v$ from the diffraction peak center in the
observation plane. In a thought single slit experiment with
the same experimental parameter as in the experiment
referred to above, the node should have been found by the
ionization detector about 63 $\mu$m from the diffraction
pattern center. In the real experiment, the first node
(indicating orthogonality) was found 12 $\mu$m from the
central diffraction peak. Since $\tsuper$ scales
proportionally to this distance, we can compute the
interferometric macroscopality of the superposition state
as $ M= 63/12 \approx 5.2$. This is perhaps smaller than
one would expect.

In a later refinement \cite{Nairzetal2003}, the molecular
beam was velocity filtered around 110 ms$^{-1}$. In this
case the the coherence length of the molecules increased
significantly, resulting in several diffraction fringes.
However, the interferometric macroscopality of the state
was equal to that in the earlier experiment. With a single
slit, the first diffraction node would have been expected
at 125 $\mu$m from the diffraction center. With the grating
the first node is found at 24 $\mu$m from the center. Thus,
$M=125/24 \approx 5.2$.

From the considerations above, we see that the path towards
larger interferometric macroscopality lies not in employing
molecules with larger mass, but only in making the relative
slit separation greater. 

It is of course an impressive fact to produce a matter
wave consisting of relatively heavy molecules with a
${}>100\mathrm{ nm}$ transverse coherence length. However,
a matter wave is not automatically the same as a
macroscopic superposition state in the sense of
Schr\"odinger.

\subsection{SQUID interference}
\label{subsec. SQUID}

Another physical system where macroscopic superposition
states have been created and measured, is superconducting
interference devices (SQUIDs). These devices consist of a
superconducting wire loop incorporating Josephson junctions
\cite{Friedmanetal2000,Vanderwaletal2000}. In these
junctions, the magnetic flux through the ring is quantized
in units of the flux quantum $\phi_0 = h/2 e$, where $e$ is
the unit charge. The supercurrent in the loop can be
controlled by applying an external magnetic field. When the
external field magnetic flux $\phi_x$ through the ring is
about one half a flux quanta, the supercurrent in the ring
can either flow in such a way that the induced magnetic
flux cancels $\phi_x$ or so that it augments it. The
corresponding ``fluxoid'' states correspond to zero and one
flux quanta respectively. The SQUID potential $U$ is given
of the sum between the magnetic energy of the ring and the
Josephson coupling energy of the junction(s): \beq U =
\frac{1}{2} \left [ \frac{\phi_0^2}{ L}\left (
\frac{\phi-\phi_x}{\phi_0} \right )^2 - \frac{\phi_0
I_c}{\pi}\cos(2 \pi \phi/\phi_0)\right ] ,\label{eq:
Josephson}\eeq where $L$ is the the ring inductance and
$I_c$ is the junction critical current. By tuning the
induced flux, a double-well potential as a function of the
flux $\phi$ that threads the ring can be created. To a
first approximation, each well can be approximated by a
harmonic potential with (approximately) equidistantly
spaced energy states. However, there is a finite barrier
between the wells, and this barrier can be tuned with the
help of the applied magnetic field. For certain values of
the external field, an excited level of each of the wells
line up, and tunnelling between the states is possible. In
one of the experiments \cite{Friedmanetal2000}, the
tunnelling transition probability is monitored as a
function of the applied magnetic flux and the frequency of
a microwave frequency pulse. Under certain conditions it is
possible to create an equal superposition of the fluxoid
states $\ket{0}$ and $\ket{1}$. The odd and even
superposition states differ in energy by about 0.86 $\mu$eV
($\Delta E/k_B \approx 0.1$ K, where $k_B$ is Boltzmann's
constant).

To analyze the fluxoid superposition state, the preferred
operator is the magnetic flux through the ring. The
difference in the (mean) magnetic flux of the two bare
fluxoid states $\ket{0}$ and $\ket{1}$ is deduced to be
about $\phi_0/4$. In order to estimate the interferometric
macroscopality of this state we need to estimate the flux
interaction needed to evolve a ``classical'' flux state
into an orthogonal one. Since the considered fluxoid states
are discrete, they nominally do not evolve under the action
of the flux operator (only their overall phase evolves). To
make an estimate of how rapid the evolution of a
``classical'' flux state would be, we use the fact that the
SQUID's potential wells are approximately harmonic, and
that classically, the state would be confined to only one
of the wells. We can then construct a coherent flux state
confined to one of the wells.

To estimate the state's dispersion (width) when expressed
in the flux operator, we note that the potential well level
spacing of the SQUID $\Delta {\cal E}$ in each potential
well is about is about 86 $\mu$eV ($\Delta {\cal E}/k_B
\approx 1$ K). The numerical values for the potential are
$\phi_0^2/2 L \approx 55.6$ meV ($\phi_0^2/2 L k_B \approx
645$ K), and $\phi_0 I_c/2 \pi \approx 6.5$ meV ($\phi_0
I_c/2 \pi k_B\approx 76$ K). In a (mechanical) harmonic
oscillator, we have $U=\kappa x^2/2$, an energy level
spacing of $\hbar \omega = \hbar \sqrt{\kappa/m}$, and a
position dispersion of \beq \Delta x \approx
\frac{1}{2}\sqrt{\frac{\hbar}{\omega m}} =
\frac{1}{2}\sqrt{\frac{\hbar}{\sqrt{\kappa m}}} ,
\label{eq: dispersion}\eeq 
where $\kappa$ is the spring constant and $m$ is the
oscillator mass. Taylor expanding the potential in
(\ref{eq: Josephson}) around the point $\phi_x=0$ and using
the analogy with the mechanical harmonic oscillator, we
find that the flux dispersion of the flux coherent state is
\beq\Delta \phi \approx \frac{\phi_0}{2}
\sqrt{\frac{1}{1290 + 4 \pi^2 76}} \approx 7.6 \cdot
10^{-3} \phi_0. \eeq Hence, the state's interferometric
macroscopality is \beq M = \frac{\phi_0 /4}{7.6 \cdot
10^{-3} \phi_0} \approx 33 . \eeq This interferometric
macroscopality is impressive, but significantly different
from what one might naively guess, considering that the
flux difference between the states $\ket{0}$ and $\ket{1}$
corresponds to a local magnetic moment of about
$10^{10}$~$\mu_B$ \cite{Friedmanetal2000}.

\subsection{Quantum superpositions of a mirror}

In the literature it has been proposed that emerging
technology will soon allow one to put a mirror into a
superposition of positions \cite{Marshalletal2002}. The
idea is to suspend a small ($10 \times 10 \times 10$
$\mu$m) mirror at the tip of a high Q-value cantilever. The
mirror would form the end-mirror of a plano-concave, high-Q
cavity. This cavity would form one arm of a Michelson
interferometer. In the other arm there would be a rigid
cavity with equal resonance frequency and finesse. When a
single photon is incident on the Michelson interferometer
input port, with probability $1/2$, the photon would either
be found in the rigid cavity or in the cavity with the
cantilever suspended mirror. The photon pressure would
shift the position of the suspended mirror, and relatively
quickly, the position of the cantilever and the photon
state would be entangled. The authors proceed to analyze
how well this entanglement could be detected by detecting
in what port the photon exits the Michelson interferometer
as a function of time. Quite intuitively, if the photon
exerts pressure on the mirror for a whole oscillation
period of the mirror, the work exerted by the photon on the
mirror when the two are moving codirectionally, is
cancelled by the work the mirror exerts on the photon mode
when he mirror and photon move contradirectionally. Hence,
there will be a revival in the photon interference
visibility after an interaction time equal to the mirror
oscillation period.

Of course, dissipation (decoherence) will impede one to
observe this revival, but the authors deduce that in a cool
enough environment, it is possible to find somewhat
realistic parameters allowing the quantum superposition to
be detected through the photon's visibility revival. The
authors suggest that under these conditions, the separation
between the two mirror positions in the superposition
correspond to the width (dispersion) of a coherent state
wavepacket. This immediately lets us deduce that the
interferometric macroscopality of the suggested experiment
is of the order unity, in spite of involving an astronomic
number of atoms, $\sim 10^{14}$.

\section{Conclusions}

In this paper, a measure has been presented to quantify how
``macroscopic'' some superpositions realized in different
experiments are. The measure is based on the utility of the
superposition state as a probe in interference experiments,
quantified by the difference in time or, more generally, in
interaction strength needed to make a macroscopic
superposition or, respectively, the single macroscopic
states evolve into orthogonal states by means of a unitary
transformation generated by a particular preferred
observable.

The proposed measure gives values for the ``interferometric
macroscopalities'' of recent experiments that are perhaps
smaller than expected; this is due to the fact that the
measure is not directly related to the number of particles,
or excitations, involved in the experiment, but rather to
the interference properties deriving from large separations
of two superposed states as measured by the preferred
observable.

Let us finally remark again that the issue about the word
`macroscopic' is very subjective, but it is interesting as
well, and can provide some insight in the way we look and
use quantum and non-quantum mechanical concepts.

\begin{acknowledgments}
This work was inspired by a talk given by Dr.\ 
H.~Takayanagi, NTT Basic Research Laboratories, at the
QNANO '02 workshop at Yokohama, Japan. The work was
supported by the Swedish Research Council (VR).
\end{acknowledgments}



\end{document}